\documentclass[11pt]{article}

\usepackage{amsmath,amsfonts,amssymb,amsthm,bbm,bm}
\usepackage{pdfsync}
\usepackage{graphicx}
\usepackage[latin1]{inputenc}
\usepackage{hyperref}
\usepackage[all]{xy}
\usepackage{stmaryrd}
\usepackage{rotating}

\newcommand{\bit}{\begin{itemize}}
\newcommand{\eit}{\end{itemize}}
\newcommand{\bd}{\begin{description}}
\newcommand{\ed}{\end{description}}

\newcommand{\bc}{\begin{center}}
\newcommand{\ec}{\end{center}}
\newcommand{\Ref}[1]{(\ref{#1})}

\newcommand{\C}{{\mathbb C}}

\newcommand{\scr}{\scriptscriptstyle\rm}

\newcommand{\su}{{\mathfrak{su}}}

\renewcommand{\sl}{{\mathfrak{sl}}}

\newcommand{\be}{\begin{equation}}
\newcommand{\ee}{\end{equation}}
\newcommand{\bea}{\begin{eqnarray}}
\newcommand{\eea}{\end{eqnarray}}
\newcommand{\bs}{\begin{subequations}}
\newcommand{\es}{\end{subequations}}
\newcommand{\nn}{\nonumber}

\newcommand{\w}{\wedge}

\newcommand{\f}{\frac}

\def\p{\partial}

\newcommand{\re}{\mathrm{Re}}
\newcommand{\im}{\mathrm{Im}}

\newcommand{\na}{\nabla}

\def\a{\alpha}
\def\b{\beta}
\def\g{\gamma}
\def\d{\delta}
\def\eps{\epsilon}

\def\th{\theta}

\def\k{\kappa}
\def\l{\lambda}
\def\m{\mu}
\def\n{\nu}

\def\r{\rho}

\def\s{\sigma}
\def\t{\tau}

\def\z{\zeta }

\def\om{\omega}
\def\G{\Gamma}
\def\D{\Delta}

\newcommand{\og}[1]{\overset{\scriptscriptstyle g}{#1}{}}

\newcommand{\nt}{{- \hspace{-5.5pt}2}}

\begin{document}

\title{\bf Raychaudhuri and optical equations \\ for null geodesic congruences with torsion } 

\author{\Large{Simone Speziale$^1$}
\smallskip \\ \small{$^1$ Aix Marseille Univ., Univ. de Toulon, CNRS, CPT, UMR 7332, 13288 Marseille, France} }
\date{\today}

\maketitle

\begin{abstract}
\noindent 
We study null geodesic congruences (NGCs) in the presence of spacetime torsion, recovering and extending results in the literature. Only the highest spin irreducible component of torsion gives a proper acceleration with respect to metric NGCs, but at the same time obstructs abreastness of the geodesics. 
This means that it is necessary to follow the evolution of the drift term in the optical equations, and not just shear, twist and expansion. We show how the optical equations depend on the non-Riemannian components of the curvature, and how they reduce to the metric ones when the highest spin component of torsion vanishes.
\end{abstract}

\tableofcontents

\section{Introduction}

Torsion plays an intriguing role in approaches to gravity where the connection is given an independent status with respect to the metric. This happens for instance in the first-order Palatini and in the Einstein-Cartan versions of general relativity (see \cite{Hehl:1976kj,Hehl:1994ue,Shapiro} for reviews and references therein), and in more elaborated theories with extra gravitational degrees of freedom like the Poincar\'e gauge theory of gravity, see e.g.  \cite{Tseytlin:1981nu}. 
The presence of torsion modifies the geodesic and geodesic deviation equations, so if the metric is invertible, one has two notions of geodesics in spacetime: the metric ones, defined by the Levi-Civita connection and which extremize the path's length; and the torsional ones, given by the full connection and autoparallel with respect to it. In general relativity, (time-like and null) geodesics play a constructive role as the trajectories followed by test particles. 
The physical relevance of torsional geodesics is on the other hand unclear: unlike for the metric one, they do not arise from the test particle approximation of the energy momentum tensor conservation law, see \cite{Yasskin:1980bu,Puetzfeld:2007ye,Hehl:2013qga} for results and discussions. 
Furthermore, the example of the Papapetrou equation shows that spinning test matter, the simplest candidate as a source of torsion, does not follow torsion-full geodesics.
 
In spite of these limitations, the geodesic deviation and associated Raychaudhuri equations with torsion have been studied in the literature, often motivated by applications to modified theories of gravity, see e.g. \cite{Hehl:1976kj,Griffiths:1981ym} for early work and more recently \cite{Luz:2017ldh,Dey:2017fld,Puetzfeld:2018cnf,Akhshabi:2018cvb}. In this brief note we restrict attention to null geodesic congruences (NGCs), rederive results of \cite{Luz:2017ldh} and extend the analysis to include the null Raychaudhuri and the rest of Sachs' optical equations with torsion. We do so for a completely arbitrary torsion, without specifying an action principle or matter coupling.

The main technical difficulty when studying geodesics with torsion is that the orthogonality of a Lie-dragged connecting vector is not preserved in general. 
Hence, one cannot restrict attention to a bundle of `abreast' null geodesics, 
to use the terminology of \cite{PenroseRindler2}, as it is customary in the Riemannian case. This introduces the need to follow not only the expansion, shear and twist, but also a drift term, corresponding to two non-orthogonal components of the displacement tensor.
Furthermore, the drift term is not frame-invariant already in the metric case, meaning it depends explicitly on the choice of transverse vector used to define the null congruence's geometric quantities. In the presence of torsion the situation is worse: also shear, twist and expansion are not frame-invariant, since the displacement tensor is given not just by the usual covariant gradient of the geodetic vector field, but also by  frame-dependent torsion components. Therefore, different local Lorentz observers will disagree on the transverse distance between the rays of the NGC,
as opposed to what happens for the set for abreast metric geodesics.
The optical equations we derive are however frame-invariant, even though the individual geometric quantities are not.

In spite of these limitations, computing the optical equations in the presence of torsion is a simple exercise carried out best with the use of the Newman-Penrose formalism, and has the nice pay-off of allowing one to review some usually marginal aspects of metric NGCs, as well as technical properties of the curvature tensor in the presence of torsion. An interesting aspect of the optical equations in the presence of torsion is that they depend also on the irreducible components of the curvature that are absent in the Riemannian case, like the antisymmetric part of the Weyl and Ricci tensors. 
This dependence however cancels out in the Raychaudhuri equation.

As we show here, the obstruction to abreastness comes only from the 
spin-2 irreducible component of torsion. For the most common framework of fermions minimally or almost-minimally coupled to the first order Einstein-Cartan action, there is no spin-2 part, and one can work with an abreast bundle, and furthermore its shear, twist and expansion are frame-invariant, as in the Riemannian case. The torsional geodesics coincide in this case  with the metric ones up to a difference in inaffinity determined by the vector (with spin 1 and spin 0 components) trace-part of torsion. Accordingly, also the Raychaudhuri equation coincides with the metric one, a result that was used in \cite{DeLorenzo:2018odq}, and so do the optical equations for the shear and twist (up to gauge choice on the space-like dyad used). 
The drift equation on the other hand always differs, because the notion of drift depends  on the choice of transverse vector, and its evolution is non geodetic and feels even a completely antisymmetric torsion.

We use spacetime metric with mostly plus signature, which means that we have to reverse the sign in the definitions of the NP scalars (see e.g. Appendix of \cite{Ashtekar:2000hw}), in order to maintain use of the various field equations and identities. The complete list, together with geometric interpretations, is reported in Appendix~\ref{AppNP} for convenience of the reader.
For the notation, we use $\na,\G$ for a generic connection, and $\og\na,\og\G$ when referring to the Levi-Civita one. The spin coefficients and the other NP scalars refer always to an arbitrary connection, so we extend their use in the same way: e.g. $\s$, $\Psi_2$ will refer to a torsion-full coefficient, and $\og\s$, $\og\Psi_2$ to the restriction to its Levi-Civita part. This with the exception of Section 2: since in that Section we only review Levi-Civita quantities, we avoid putting the superscript everywhere for ease of reading.

\section{Metric null geodesic congruences and optical equations}
In this preliminary Section, we review familiar and less familiar aspects of metric NGCs, in particular the reason and interest of considering only abreast rays, and the derivation of Sachs' optical equations using the NP formalism. 
All derivatives, connections and curvature terms appearing in this Section are understood to be given with respect to the Levi-Civita connection without additional decorations,
to avoid making the equations look unnecessarily like an italian baroque church.

\subsection{Null geodesic congruences and kinematical quantities}

We denote by $l^\m$ a null, geodesic vector field, not necessarily affinely parametrized, 
\be\label{NGCg}
l^2=0, \qquad   D l^\m =   k l^\m, \qquad   D:=l^\n \na_\n.
\ee
To study the null geodesic congruence (NGC) generated by $l$ one first introduces a transverse null vector $n$ such that $l_\m n^\m = -1$, and defines the projector
\be\label{defperp}
 \bot_{\m\n}:=g_{\m\n} +2l_{(\m}n_{\n)}
\ee
on 2d space-like surfaces $S$. It is convenient to introduce also a (complex) dyad $(m^\m,\bar m^\m)$ such that 
\be
\bot_{\m\n}=2m^{(\m}\bar m^{\n)}, \qquad g_{\m\n}=-l_{(\m}n_{\n)}+m_{(\m}\bar m_{\n)}.
\ee 
The doubly-null tetrad $(l,n,m,\bar m)$ 
allows us to use the Newman-Penrose (NP) formalism. All components of the connection are represented by (complex) spin coefficients labelled by a greek letter, and endowed with a specific geometric interpretation. For convenience of the reader unfamiliar with the NP formalism, we summarize definitions and geometric properties in Appendix~\ref{AppNP}, referring to the monographs \cite{PenroseRindler2,Chandra} for more details.

Two technical remarks are useful at this point: first, the 2d surfaces identified by $n$ are in general not integrable. From the Lie bracket
\be\label{mmbar}
[m,\bar m]^\m =(\m-\bar\m) l^\m+(\r-\bar\r)n^\m-(\a-\bar\b) m^\m+(\bar\a-\b)\bar m^\m,
\ee
we see that their integrability requires $\im(\r)=0=\im(\m)$, namely the vanishing of the twist of the $l^\m$ and $n^\m$ congruences. 
Second, the transverse vector $n$ is not unique. There is a 2-parameter family of choices, 
corresponding to
$l$-preserving Lorentz transformations of the doubly-null tetrad $(l,n,m,\bar m)$ (called class $I$ transformation in the nomenclature of \cite{Chandra}), given by
\be\label{classI}
n^\m\mapsto n^\m+\bar a m^\m +a\bar m^\m+|a|^2 l^\m,\qquad m^\m\mapsto m^\m+a l^\m, \qquad a\in\C.
\ee
If $l$ is hypersurface orthogonal (namely its twist $\im(\r)$ vanishes, since it is null and geodesic), then the gauge freedom \Ref{classI} can be used to achieve $\im(\m)=0$, in which case $(m,\bar m)$ are integrable vector fields and span the 2d surface of generators of null geodesics ruling the hypersurface of $l$. In general, a convenient gauge choice is to require 
$m$ (and thus $\bar m)$ to be parallel transported along $l$, namely (see Appendix~\ref{AppNP})
\be\label{gaugeDm}
D m^\m=0 \quad \Leftrightarrow \quad \pi=0=\im(\eps). 
\ee
This can be achieved using first \Ref{classI} to set $\pi=0$, and then the freedom of O(2) rotations in the $(m,\bar m)$ plane (class $III$ transformations) to set $\im(\eps)=0$. It also implies that $D n^\m=-k n^\m$.\footnote{Which doesn't mean that $n$ is geodesic! But only parallel transported along $l$. The (orthogonal) acceleration of $n$ is measured by the spin coefficient $\n$, see Appendix~\ref{AppNP}.} We will consider this gauge further below.

To study the geodesic deviation, one introduces a connecting vector $\eta$  Lie dragged by $l$,
\be\label{Lieeta}
\pounds_l \eta^\m = 0,
\ee
as to have local coordinates defined by $l$ and $\eta$ forming a grid (or equivalently, that we have a smooth 1-parameter congruence of geodesics connected by $\eta$). 
Thanks to this condition, the displacement of $\eta$ along $l$ is $\eta$-independent, 
\begin{align}\label{Dgeta}
&   D \eta_\m =   B_{\m\n}\eta^\n, \qquad   B_{\m\n}:=  \na_\n l_\m,
\end{align}
and the displacement tensor $B$ satisfies the properties
\be
 l^\m   B_{\m\n}=0, \qquad l^\n   B_{\m\n}=  k l_\n. \label{Bg}
\ee
Using \Ref{defperp} we project $B$ on the surface $S$, and decompose it in irreducible representations
\be
B^\bot_{\m\n}:=\bot_{\m\r}\bot_{\n\s} B^{\r\s} = \s_{\m\n}+\om_{\m\n}+\f12\bot_{\m\n}\th,
\qquad 
\ee
where the symmetric-traceless, antisymmetric and trace parts are given respectively by
\be\label{perpBg}
\s_{\m\n}:= (\bot_{(\m}{}^\r \bot_{\n)}{}^\s-\f12\bot_{\m\n}\bot^{\r\s})  B_{\r\s},
\qquad   \om_{\m\n}:= \bot_{[\m}{}^\r \bot_{\n]}{}^\s B_{\r\s},  \qquad     \th:= \bot^{\m\n}  B_{\m\n}.
\ee
These three quantities are captured by the two spin coefficients
\begin{subequations}\label{rsg}\begin{align}
&  \s := -m^\m m^\n  \na_\n l_\m = -m^\m m^\n  \s_{\m\n}, \\
&  \r := -m^\m \bar m^\n  \na_\n l_\m = -\f12 \th-m^\m \bar m^\n  \om_{\m\n}.
\end{align}\end{subequations}
We also recall for later use that 
\be\label{B2g}
B^\bot_{\m\n} B^\bot{}^{\n\m} = \s_{\m\n}\s^{\m\n} - \om_{\m\n}\om^{\m\n} +\f12\th^2 = \r^2+\bar\r^2+2|\s|^2.
\ee

A word about the frame-invariance of these quantities: under a class $I$ transformation we have
\be\label{rsI} 
\r\stackrel{I}{\mapsto}\r+\bar a\k, \quad \s\stackrel{I}{\mapsto} \s+a\k, 
\ee
where $\k := -m^\m Dl_\m=0$ for a geodesic vector field. Hence, the scalar description  \Ref{rsg} 
is frame-independent for geodesic congruences.
This is one of the numerous advantages of working with the NP formalism instead of tensors, since $\s_{\m\n}$ and $\om_{\m\n}$ in \Ref{perpBg} are not frame-independent. Only their squares or the further projections along the complex dyad are. The squares are the only quantities entering the Raychaudhuri equation, making it frame-independent.

These three quantities have a precise geometric meaning, respectively in terms of the shear, twist and expansion of the congruence. 
It is immediate to see that the expansion $\th$ measures the variation of the area element of the  2d space-like surfaces, since a standard calculation gives
\be
\th =-2\re(\r) =
\na_\m l^\m -k  = -\f12\perp_{\m\n}\!\pounds_l \!\perp^{\m\n}  =: \pounds_l \ln\sqrt\g, \label{thg}
\ee
where in the last equality we introduced a shorthand notation to remind us that if we take adapted coordinates to the NGC (e.g. Bondi coordinates), then the projector only has transverse components, and we denoted by $\g$ its $2\times 2$ non-zero determinant. 
To visualize the geometric meaning of $\s$ and $\im(\r)$,
we follow \cite{PenroseRindler2} and parametrize the connecting vector in terms of the doubly-null tetrad,
\be\label{etapar}
\eta^\m = -g l^\m - h n^\m +\bar \z m^\m +\z \bar m^\m. 
\ee
Inserting this decomposition on both sides of \Ref{Dgeta}, and projecting along the basis components, one derives the propagating equations
\begin{subequations}\label{Dgdec1}\begin{align}
&   D g = \g h +(\pi-\a-\bar\b)\z +{\rm cc.} \\
&   D h = k h \label{Dgh}\\
&   D \z =  (\t+\bar\pi) h-( \r +\bar\eps-\eps)\z- \s\bar\z \label{Dgz} 
\end{align}\end{subequations}
where we used $k=2\re(\eps)$, see Appendix~\ref{AppNP}. The quantity $g$ and its equation are of little interest:
even once we have entirely fixed our $(l,n,m,\bar m)$ frame, the restriction to a Lie-dragged connecting vector still leaves the freedom to change $\eta^\m\mapsto \eta^\m+b l^\m$ with $\pounds_l b=0$, thus making the function $g$ largely irrelevant. 

In the second equation, $h=\eta_\m l^\m$ measures the non-orthogonality of the connecting vector with respect to $l$. Its evolution \Ref{Dgh} (which can also be immediately derived from the Lie-dragging of $\eta$) implies that $h=0$ is preserved along the NGC. The set of null geodesics in the congruence related by an orthogonal connecting vector are called `abreast',\footnote{Namely one next to the other. This can be most easily visualized if the twist vanishes, then the abreast null geodesics are those lying in the same hypersurface.}
and play a privileged role in the study of the NGC.
In fact we see from \Ref{Dgz}
that the deformation of a bundle of rays with $h=0$  is self-contained in the complex $\z$ plane, and it is easy to see writing $\z$ in polar decomposition that  $\s$ produces a shear of the bundle, $\re(\r)$ a contraction or expansion depending on its sign,\footnote{The geometric relevance of the $\z$ plane can be completed recovering the interpretation of the expansion already given above. Following again \cite{PenroseRindler2}, we consider a small triangle in the $\z$ plane, identified say by the origin and two points $\z_1$ and $\z_2$. Its area is given by 
$
A_t:=\f i2 (\z_1\bar\z_2-\z_2\bar\z_1),
$
and \Ref{Dgz} for $D m^\m=0$ gives
\be\nn
D A_t = -2\re(\r) A_t -h_1\im(\t\bar\z_2)+h_2\im(\t\bar\z_1).
\ee
Hence if $h=0$ (and only if) the (logarithmic) variation of the triangle area is given by  the expansion, $
\th =  D \ln A_t.
$
}
whereas $\im(\r)-2\im(\eps)$ a twist.
The twist introduced by $\im(\eps)$ is due to the rotation of the complex dyad while propagated along the NGC and thus a gauge artifact. Choosing a parallel propagated complex dyad \Ref{gaugeDm} the equation for abreast rays reduces to the more familiar form
\be
D \z =-\r \z- \s\bar\z.
\ee

For non-abreast geodesics, one has also a drift term measured by $\t+\bar\pi$ -- or $\t$ alone in the gauge \Ref{gaugeDm}.
As a side remark, we notice that this quantity coincides with the non-integrability of the time-like planes spanned by $l$ and $n$,
\be
m_\m[l,n]^\m=\t+\bar\pi.
\ee
Hence, although it drops out from the optical equations if one restricts to orthogonal connecting vectors, it plays an important dynamical role when the full set of Einstein's equations is considered, since it is one of Sachs' constraint-free data at the 2d corner between two null hypersurfaces \cite{Sachs62} (see also \cite{IoElena} and references therein).

Apart from the simplicity of not having a drift term, there is a related but more fundamental property
of abreast geodesics which is worth recalling.
Under the change of frame \Ref{classI} we have
\be\label{zetaFD}
h\stackrel{I}\mapsto h, \quad \z\stackrel I\mapsto \z-ha,
\ee
hence for abreast rays the function $\z$ is frame-independent. In fact, since $\eta^2=-2gh+|\z|^2$, 
the transverse distances $|\z|^2$ of abreast rays are invariant under all local Lorentz transformations. 
We thus have a stronger frame-independent property: all local observers agree on the transverse distances among abreast rays.
This property is spoiled for non-abreast geodesics, because of \Ref{zetaFD}, and further notice that although $\r$ and $\s$ are frame-invariant, the drift term is not:
\be
\t \stackrel{I}{\mapsto} \t+a\r+\bar a\s+|a|^2 \k,
\ee
which is not preserved even for geodesics.

As a final comment on the drift, we notice that it is given by two non-orthogonal components of the displacement tensor $B$, 
\be\label{B||}
\t:=-m^\m \D l_\m = -m^\m n^\n B_{\m\n},
\ee
showing explicitly the statement that for abreast bundles all information is carried by the orthogonal part of $B$.

\subsection{Dynamics: Raychaudhuri and optical equations}
If one is interested in the Raychaudhuri equation alone, the NP formalims is largely unnecessary, and it is customary to derive it using tensors.
One computes first
\be\label{DBg}
l^\r \na_\r (g^{\m\n}  B_{\m\n}) = -  R_{\m\n}l^\m l^\n -   B_{\m\n}   B^{\n\m} + \na_\m(  k l^\m)
\ee
from the commutator of two covariant derivatives.
Using then $g^{\m\n}B_{\m\n} = \na_\m l^\m = \th+k$ and 
\be
  B^\bot_{\m\n}   B^{\bot\n\m} =   B_{\m\n}   B^{\n\m} -  k^2 
\ee
which follows from \Ref{Bg},
one immediately arrives at the familiar Raychaudhuri equation,
\be\label{Dthg}
  D \th =  -\f12 \th^2- \s_{\m\n}^2+ \om_{\m\n}^2 -  R_{\m\n}l^\m l^\n+   k \th.
\ee

The NP formalism becomes on the other hand very convenient to go beyond this equation and study the evolution of shear and twist as well.
To that end, 
we need first the geodesic deviation equation. Acting with $D$ on \Ref{Dgeta} one gets
\begin{align}
D^2 \eta^\m  = R^\m{}_{\l\r\n}l^\l l^\r\eta^\n +\eta^\n\na_\n(k l^\m).
\end{align}
As before, we use the parametrization \Ref{etapar} and project this vectorial equation along the basis components. 
To simplify the equations without loss of geometric information, we choose from now the partial internal gauge \Ref{gaugeDm} as to have the complex dyad parallel propagated along the NGC, a customary choice in both the NP \cite{PenroseRindler2} and tensorial \cite{Wald} derivation of the optical equations. One then finds
\begin{subequations}\label{Dgdec2}\begin{align}
&   D^2 h = k(2Dh+hD\ln k-hk), \\
&   D^2 \z 
= ( \Psi_1+ {\Phi}_{01}) h -  \Phi_{00}\z- \Psi_0\bar\z  +   k(  \t h -  \r \z -  \s \bar \z),\label{D2zg}
\end{align}\end{subequations}
where we used 
\be\label{metanal}
m_\m\eta^\n\na_\n l^\m =\t h -\r\z-\s\bar\z
\ee
and $\Psi$ and $\Phi$ are components respectively of the Weyl and Ricci tensors, see Appendix~\ref{AppNP} for definitions.\footnote{For completeness, we report also the equation for $g$:  \begin{align*}
D^2 g &= \Big(2\re(\Psi_2)+2\Phi_{11}-\f1{12}R \Big)h -2\re\Big((\bar\Psi_1+\Phi_{10})\z\Big) 
+(g  D+h \D - \bar\z  \d - \z\bar{ \d})   k +   k\Big[k g + 2 \re\Big( \g h +(\a+\bar\b)\z\Big)\Big]. 
\end{align*}
}

We now substitute  \Ref{Dgdec1} into \Ref{Dgdec2} to derive relations between the spin coefficients and curvature components. The equation for $D^2h$ gives an identity, but equating the $D^2\z$ equations obtained from \Ref{Dgz} and \Ref{D2zg} one finds 
\be\label{dev2}
D(m_\m\eta^\n\na_\n l^\m) = R_{\m\n\r\s}m^\m l^\n l^\r\eta^\s + k m_\m\eta^\n\na_\n l^\m.
\ee
Next, we use \Ref{metanal} on the left-hand side, and when $D$ acts on the $(h,\z,\bar\z)$ parameters we substitute again the right-hand sides of \Ref{Dgdec1}. The result in NP language  reads
\be\label{totti}
h\Big( D  \t - \r  \t - \s\bar{ \t} -  \Psi_1 - \Phi_{01}\Big) +
\z \Big( -D  \r +  \r{}^2 + | \s|^2 +  {\Phi}_{00} +   k   \r\Big) + \bar\z \Big(-D \s + ( \r+\bar{ \r}) \s +  {\Psi}_{0} +   k   \s \Big)=0.
\ee
Requiring the equation to be satisfied for all $\eta$'s, one finds the following relations between the spin coefficients and curvature components \cite{PenroseRindler2},
\begin{subequations}\label{opt}\begin{align}
\label{S1} &   D  \r =  \r{}^2 + | \s|^2 +  {\Phi}_{00} +   k   \r, \\
\label{S2} &   D \s = ( \r+\bar{ \r}) \s +  {\Psi}_{0} +   k   \s, \\
&   D  \t =  \r  \t + \s\bar{ \t} +  \Psi_1 +  \Phi_{01}.
\end{align}\end{subequations}
These are Sachs' optical equations, here written for an arbitrary bundle of NGC with both $k$ and $h$ non-vanishing.
The set contains
\be
  D \th = -D( \r+\bar{ \r})
   = - \r{}^2- {\bar{ \r}}{}^2 - 2 | \s|^2 - 2\re( {\Phi}_{00}) +   k\th, 
\ee
which using \Ref{B2g} we recognize to be the Raychaudhuri equation in this language.

For abreast rays $h=0$, so 
the first equation in \Ref{opt} is no longer needed, and one recovers the usual basic set of Sachs' (\ref{S1}, \ref{S2}). Which in particular shows that the evolution of shear, twist and expansion is all that is needed to characterize abreast rays in the NGC.
For the non-abreast ones, one has to include the evolution of $\t$. 

Stated in other terms, projecting the Einstein's equations along a doubly-null basis $(l,n,m,\bar m)$ has the nice feature that if $l$ is geodetic, the two equations for $\r$ and $\s$ decouple from the rest, giving the optical equations describing the evolution of shear, twist and expansion of a null congruence associated with $l$. The larger system including $\t$ is also closed if $l$ is geodetic, however it depends also on $n$, and describes not just the intrinsic properties of the NGC, but also part of the dynamics of the non-orthogonal connecting vector used.

This concludes our review of the optical equations for a metric null geodesic congruence. The two possibly less familiar aspects we highlighted are:
\begin{itemize}
\item Orthogonality of the connecting vector is preserved, hence one can restrict attention to a bundle of abreast geodesics, for which the evolution is captured by shear, twist and expansion, and it is completely frame-independent, meaning independent of the choice of transverse vector $n$;
\item For non-abreast geodesics, one has to include the evolution of the drift term, which is frame-dependent.
\end{itemize}
This background will be useful to appreciate the torsion-full case, to which we now turn our attention.

\section{Curvature, torsion and their irreducible components}

In the rest of the paper, we will use $\na_\m, \G^\r_{\m\n}$ to denote a generic connection carrying torsion. 
When needed, the Levi-Civita connection or other quantities determined by the metric $g$ will be denoted by and apex, e.g. $\og{\G}^\r_{\m\n}:=\G^\r_{\m\n}(g)$. The contorsion tensor $C$ is defined by 
\begin{align}\label{GC}
\G^\r_{\m\n} =\og{\G}^\r_{\m\n}+C_{\m,}{}^\r{}_\n,
\end{align}
with the comma meant to separate the one-form index from the pair of antisymmetric fibre indices.
The torsion is most elegantly defined using the tetrad formalism by $T^I:=d_\om e^I$, and
it is related to the contorsion by
\begin{align}
& T^\r{}_{\m\n}:=e^\r_I \, T^I{}_{\m\n}(e,C) = 2C_{[\m,}{}^\r{}_{\n]}=2\G^\r_{[\m\n]}, \\
& C_{\m,\n\r} = \f12 T_{\m,\n\r} - T_{[\n,\r]\m}, \qquad C_{(\m,\n)\r}=T_{(\m,\n)\r}. \label{CofT}
\end{align}

Both torsion and contorsion transform under the $\sl(2,\C)\cong \su(2)_\C\oplus\su(2)_\C$ algebra representation
$
\bf{(\tfrac12,\tfrac12)\otimes[(1,0)\oplus(0,1)]=(\tfrac32,\tfrac12)\oplus(\tfrac12,\tfrac32)\oplus(\tfrac12,\tfrac12)\oplus(\tfrac12,\tfrac12)}.
$
This gives three irreducible components under Lorentz transformations (since the latter includes parity), see e.g. \cite{Hehl:1976kj},
\begin{align}\label{Cirreps}
& C^{\m,\n\r} = \bar{C}^{\m,\n\r} + \f23 g^{\m[\rho} \check{C}^{\n]} + \eps^{\m\n\r\s} \hat C_{\s}, \\
& g_{\m\n} \bar{C}^{\m,\n\r}=0=\eps_{\m\n\r\s} \bar{C}^{\m,\n\r}, \qquad \check C^\m:=C_{\n,}{}^{\m\n}, \qquad \hat C_\s:=\f16\eps_{\s\m\n\r}C^{\m,\n\r},
\end{align}
and identically for the torsion. The irreps are related by
\be
\bar C_{\m,\n\r} = \bar T_{\m,\n\r}, \qquad \check C^\m = \check T^\m, \qquad \hat C^\m = -\f12 \hat T_\m.
\ee
The bar used to denote the spin-2 irreps $\bar C$ and $ \bar T$ should not be at risk of confusion with complex conjugation, since these fields are real.

In the presence of torsion, the commutator of two connection gives
\be
[\na_\m,\na_\n]f^\r = R^\r{}_{\s\m\n}(\G)f^\s - T^\s{}_{\m\n}\na_\s f^\r,\label{commRiem}
\ee
and the curvature tensor $R_{\r\s\m\n}(\G)$ has 36 independent components, and not just 20 as in the metric case.
It decomposes into six irreps with the following spins,
$$
\bf{(2,0)\oplus(0,2)\oplus(1,1)\oplus(1,1)\oplus (1,0)\oplus(0,1)\oplus(0,0)\oplus(0,0)}.
$$
One can obtain the irreps of this decomposition using the original spinorial methods of \cite{PenroseRindler2} or self-dual projectors (see e.g. the identitical decomposition of the Lagrange multiplier $\phi$ in \cite{Iobimetric}). It is however simplest to use the standard decomposition, 
\be\label{Riemirreps}
R_{\m\n\r\s}(\G) = C^{\scr\G}_{\m\n\r\s} + R^{\scr\G}_{\m[\r}g_{\s]\n}- R^{\scr\G}_{\n[\r}g_{\s]\m} - \f13g_{\m[\r}g_{\s]\n} R^{\scr\G},
\ee
and recognize that it is further reducible. In particular,
\begin{align}
& C^{\scr\G}_{\m\n\r\s} = C_{\m\n\r\s} + C^{\scr A}_{\m\n\r\s} + \f1{4!}\eps_{\m\n\r\s} C^{\scr T} \ \in \ \bf{(2,0)\oplus(0,2)\oplus(1,1)\oplus(0,0)} \\ \nn
& C_{\m\n\r\s} := \f12(C^{\scr\G}_{\m\n\r\s} + C^{\scr\G}_{\r\s\m\n}) - \f1{4!}\eps_{\m\n\r\s} C^{\scr T} \\ \nn
& C^{\scr A}_{\m\n\r\s} :=\f12(C^{\scr\G}_{\m\n\r\s}-C^{\scr\G}_{\r\s\m\n}), \qquad  C^{\scr T} := -\eps^{\m\n\r\s} C_{\m\n\r\s} \\
& R^{\scr\G}_{\m\n} = R_{(\m\n)} + R^{\scr A}_{[\m\n]} \ \in \ \bf{(1,1)\oplus(0,0)\oplus(1,0)\oplus(0,1)}
\end{align}

We keep the NP notation for the complex scalars built out of the Weyl-like (`like', because it is not purely Riemannian but depends on torsion as well) tensor $C_{\m\n\r\s}$ and Ricci-like $R_{(\m\n)}$, see Appendix~\ref{AppNP}. We refrain from introducing an NP notation for the non-Riemannian parts, since they will play a limited role in this short note, although this is something interesting to explore, if it has not yet been done in the literature.\footnote{One example we are aware of is 
\cite{griffiths1982spin}, but the notation there proposed for torsion simply mimics the one for the spin connection and misses the irrep decomposition \Ref{Cirreps}, making it not particularly efficient. It further seems to miss the irrep $C^{\scr T}$ of the curvature, which is possibly inadvertently included in $\Psi_2$.}

We will on the other hand often abridge the scalar products as (complex) components of the tensors, e.g. $R_{lnlm}:=R_{\m\n\r\s}l^\m n^\n l^\r m^\s$.

Finally, we recall that using \Ref{GC},
\begin{align}\label{RiemmG}
R_{\m\n\r\s}(\G) &= R_{\m\n\r \s}(e)+ 2\nabla_{[\r}C_{\s],\m\n} + 2C_{[\r,}{}^{\l}{}_{\s]} C_{\l,\m\n} - 2C_{[\r|,\m\l}C_{\s],}{}^\l{}_{\n} \\
&= R_{\m\n\r \s}(e)+ 2\og\nabla_{[\r}C_{\s],\m\n}  + 2C_{[\r|,\m\l}C_{\s],}{}^\l{}_{\n}, \\
R_{\m\n}(\G) &= R_{\m\n}(e)+ \nabla_{\n}C^{\s}{}_{\m\s} -\na_\s C_{\n,\m}{}^\s - C_{\n,\m\l}C_{\s}{}^{\l\s} + C_{\s,\l\n}C^{\l,\s}{}_{\m}, \label{RicciGC}
\end{align}
so contorsion enters both the Riemannian $\bf{(2,0)\oplus(0,2)\oplus(1,1)\oplus(0,0)}$ and non-Riemannian $\bf{(1,0)\oplus(0,1)\oplus(1,1)\oplus(0,0)}$ components.

\section{Torsion-full null geodesic congruences}

We consider a null geodesic vector field $l^\m$, not necessarily affinely parametrized,
\be\label{NGC}
l^2=0, \qquad D l^\m = k l^\m, \qquad D:=l^\n\na_\n.
\ee
This is the same set-up as before, except that covariant derivatives now carry torsion, therefore the trajectory and the inaffinity differ from the metric case.\footnote{Since the connection without the metric defines an affine structure, the torsion-full geodesics could also be called affine geodesics. This would be however an unfortunate choice for the guaranteed risk of confusion with an affinely parametrized geodesic, therefore we will avoid it and always specify that we are referring to torsion-full geodesics.}
To expose the difference we use \Ref{GC},
\be
D l_\m = \og D l_\m +C_{\n,\m\r} l^\n l^\r = \og D l_\m +T_{\n,\m\r} l^\n l^\r.
\ee
If 
$l$ is metric-geodetic and
torsion is aligned with it in the following sense, %
\be\label{Cll}
C_{\n,\m\r}l^\n l^\r = T_{\n,\m\r}l^\n l^\r = c \, l_\m, \qquad c=T_{lln},
\ee
then the torsion-full geodesics collapse on top of the metric ones, up to an inaffinity
\be
k \stackrel{\Ref{Cll}} = \og k+c.
\ee

To visualize the meaning of this condition, we use the decomposition \Ref{Cirreps}, which gives
\be
T_{\n,\m\r}l^\n l^\r = \bar T_{\n,\m\r}l^\n l^\r -\f13 l_\m \check T_\n l^\n = c \, l_\m.
\ee 
In particular, 
\be\label{Special}
\bar T_{llm} = 0, \qquad c =T_{lln}=\bar T_{lln}-\f13\check T_\n l^\n,
\ee
or the same equations with the contorsion $C$.
This shows that \Ref{Cll} is a restriction only on the spin-2 part of torsion $\bar T$.
We thus recover the well-known fact that geodesics are unchanged by a completely antisymmetric torsion, and observe that the trace part $\check T$ introduces only an inaffinity acceleration. It is only the spin-2 part that introduces a proper (i.e. orthogonal) acceleration modifying the trajectory of the metric geodesics. We can draw a qualitative analogy with the Riemann tensor, whose most non-trivial geometric content is carried by the  highest spin component, the Weyl tensor.\footnote{This is different for time-like geodesics, where also the trace part $\check T$ contributes to a proper acceleration. We also point out that the most commonly used source of torsion, fermions in a minimal or almost-minimal coupling (see e.g. \cite{Alexandrov:2008iy,IoDario}) only generate vector $\check T$ and axial vector $\hat T$ torsion, namely spins 1 and 0.}

A special case is when the spin-2 part of torsion completely vanishes. In this case \Ref{Cll} is satisfied for \emph{any} null vector $l$,
with trace-part $\check T$ and completely antisymmetric part $\hat T$ left arbitrary, and we identify
\be\label{cnt}
c \stackrel\nt = - \f13 \check T_\m l^\m  = - \f13 \check C_\m l^\m.
\ee
If one further requires $c=0$ for all $l$, namely the complete matching of all metric and torsion-full geodesics including the inaffinity, then torsion must be completely antisymmetric.
Here and in the following the symbol ${{- \hspace{-7pt}2}}$ means $\bar T=0$. 

The above considerations mean that whatever modified Raychaudhuri and optical equations we find, they should reduce to the metric ones (at most up to a Lorentz transformation) when \Ref{Cll} holds. This will be proved explicitly below, focusing mostly on the special case $\bar T=0$. The more general aligned case \Ref{Special} leads to longer formulas without much further insight, and we will limit ourselves to reporting them explicitly for the Raychaudhuri equation.

To study the geodesic deviation equation, 
we introduce as in the metric case a connecting vector $\eta$  Lie dragged by $l$,
\be\label{Lieeta2}
\pounds_l \eta^\m = 0,
\ee
so to have local coordinates forming a grid associated 
with a congruence of geodesics. 
Notice that the Lie derivative is insensitive to torsion, and thus also this requirement.
However in the metric case this condition led to two useful properties: conservation of orthogonality of $\eta$, and identification of the displacement tensor with $B_{\m\n}:=\na_\n l_\m$. Both properties are lost in the torsion-full case. 

For the orthogonality we have:
\be
D h = l^\n \pounds_l \eta_\n + \f12\eta^\m\na_\m l^2 + T_{\m,\n\l}l^\m l^\n \eta^\l + k\, h.
\ee
The first term vanishes if we take $\eta$ Lie dragged and the second since $l$ is null everywhere. The third term however means that orthogonality is not preserved in general, but only in the special case \Ref{Cll}.

For the displacement equation we have:
\be
D \eta^\m =  \eta^\n \na_\n l_\m + \pounds_l \eta^\m  + T^\m{}_{\l\n} l^\l \eta^\n. 
\ee
The deformation of the congruence with a 
Lie-dragged $\eta$ is not measured by $B_{\m\n}:=\na_\n l_\m $ anymore, even if we are including torsion in its covariant derivatives, but by the modified tensor
\begin{align}\label{Deta}
& D \eta_\m = B'_{\m\n}\eta^\n, \\
\label{B'def}
& B'_{\m\n}: = B_{\m\n} + T_{\m,\l\n}l^\l =B_{\m\n} - 2 C_{ [\l,\n] \m} l^\l  = \og\na_\n l_\m +C_{\r,\m\n}l^\r.
\end{align}

These two reasons can motivate choosing a deformation vector that is not Lie dragged as in \Ref{Lieeta}, but rather satisfies 
\be\label{Lie2}
\pounds_l \eta^\m= - T^{\m}{}_{\l\n} l^\l \eta^\n = 2C_{[\l,\n]}{}^{\m}l^\l \eta^\n.
\ee
With this choice, orthogonality is preserved and $B_{\m\n}$ alone measures the displacement. 
However, it means that there is no coordinate grid associated with our $(l,\eta)$ frame, as one would expect for a smooth congruence, and this makes it less useful a priori. We leave further considerations on the geometric meaning of \Ref{Lie2} for future work, and keep \Ref{Lieeta2} in the following, which seems to us also supported by the coordinate analysis performed in \cite{Luz:2017ldh,Puetzfeld:2018cnf}.

\subsection{Kinematical quantities and the congruence's geometry}

We will find it useful to work with both tensors $B$ and $B'$, that as we will see have complementary properties in the presence of torsion: frame independence for the projections of $B$, and describing the geometry of the NGC for the projections of $B'$.
We begin by noticing that 
\begin{align}
& l^\m B_{\m\n} = 0, && l^\n B_{\m\n} = k l_\m, \label{lB} \\
& l^\m B'_{\m\n} = T_{\m,\l\n}l^\m l^\l, && l^\n B'_{\m\n} = k l_\m. \label{lB'}
\end{align}
We introduce as in the metric case a transverse vector $n$, and the projector \Ref{defperp}, and define the 
projected tensors $B^\bot$, $B'{}^\bot$ and their 
symmetric-traceless, antisymmetric and trace components as in \Ref{perpBg}.

We define the spin coefficients $\r$ and $\s$ as before in \Ref{rsg}, this time using $B^\bot$ which carries the torsion-full connection, as with the rest of the spin coefficients and curvature scalars.
They can be related through \Ref{B'def} to equivalent quantities $\r'$ and $\s'$ for $B'{}^\bot$ (which shouldn't be thought of as spin coefficients), as well as to their Levi-Civita correspondents for $\og B^\bot$ (namely the spin coefficients determined by the Levi-Civita connection),
\begin{subequations}\label{rsrs'}\begin{align}\label{ss'}
& \s' :=-m^\m m^\n B'_{\m\n} = \s - T_{mlm} = \og\s, \\ 
& \r' := -m^\m \bar m^\n B'_{\m\n} = \r- T_{ml\bar m} = \og\r - C_{lm\bar m}, 
\end{align}\end{subequations}
or as tensors,
\be\nn\label{B'Bg}
\th' = \og\th, \qquad \s'_{\m\n} = \og\s_{\m\n}, \qquad \om'_{\m\n}=\og\om_{\m\n} - 2m_{[\m}\bar m_{\n]} C_{\l,\r\s} l^\l m^\r \bar m^\s. 
\ee
Indulging a bit more on the traces, we have
\begin{align}
& \th' =-2\re(\r')
= \th+2T_{\m,\n\r}l^\n m^{(\m}\bar m^{\r)} 
  = \og\th=-\f12\perp_{\m\n}\!\pounds_l \!\perp^{\m\n}  =: \pounds_l \ln\sqrt\g, \label{th'}
\end{align}
with
\begin{align}\label{th}
& \th =-2\re(\r)
= \na_\m l^\m -k. 
\end{align}

Having introduced this notation, we now look at the displacement equation \Ref{Deta} projected along the basis vectors using the parametrization \Ref{etapar}, like in the metric case. We fix from now on the gauge \Ref{gaugeDm} for simplicity, now referring to the torsion-full covariant derivatives. Neglecting the irrelevant equation for $g$, we find the following propagating equations, to be compared with \Ref{Dgdec1}:
\begin{subequations}\label{Ddec1}\begin{align}
 D h &= k h + T_{\m,\r\n}l^\m l^\r \eta^\n 
 \label{Dh} 
 = (k-T_{lln})h  - T_{l l \bar m} \z - T_{l l m} \bar \z, \\ 
 D \z &= \t h- \r\z- \s\bar\z + T_{\m,\r\n}m^\m l^\r \eta^\n 
 \label{Dz}
= \t' h- \r'\z- \s'\bar\z.
\end{align}\end{subequations}
The first equation  shows that orthogonality is not preserved in the presence of generic spin-2 torsion, the key property of torsion-full geodesics discussed previously.
The second confirms that it is the components of $B'{}^\bot$ to carry the correct geometric interpretation of shear, twist and expansion, coherently with the fact that it is $B'$ that represents the true displacement tensor \Ref{B'def}; and also identifies the drift coefficient as
\be\label{tt'}
\t':=\t-T_{mln} = \og\t-C_{lmn} = \og\t + \bar{\og\pi}-\bar\pi.
\ee

On the other hand, $\r'$ and $\s'$ are not frame-independent, unlike $\r$ and $\s$. The effect 
of the class $I$ rotation \Ref{classI} preserves the $B$ projections since $\k=0$,
\be
\th\stackrel I\mapsto\th, \qquad \r\stackrel I\mapsto\r, \qquad \s\stackrel I\mapsto\s, 
\ee
but not the $B'$ ones,
\be\label{rsLorentz}
\th' \stackrel I\mapsto \th' - aT_{l\bar ml} - \bar aT_{l ml}, \qquad \r'\stackrel I\mapsto \r' - a  T_{ll\bar m}, \qquad \s'\stackrel I\mapsto \s' - a  T_{ll m}.
\ee
Given that $\th'=\og\th$ and $\s'=\og\s$, it may look surprising that these quantities are not frame-independent, like in the metric case. This is a consequence of the fact that we are following torsion-full geodesics and not metric ones, and frame-invariance of the projections depend on which of the two acceleration vanishes, $\k$ or $\og \k=\k - C_{\m,\n\r}l^\m l^\n m^r$.  For a torsion-full geodesic, $\k=0$ but not $\og \k$, hence the projections of $B^\bot$ are frame invariant but not those of $B'{}^\bot$. As for the drift, with $\k=0$ we have
\be\label{tauI}
\t' \stackrel I\mapsto \t' + a \r' +\bar a \s' - a T_{lln}. 
\ee

In summary,  one can study dynamics for the frame-invariant spin coefficients $\r$ and $\s$, but the geometric content is carried by the non-frame-invariant coefficients $\r'$ and $\s'$; and the observed non-preservability of orthogonality makes us expect that we will need to include also the drift coefficient.

\subsection*{Spin-2-less torsion}
In this subsection we present  the formulas for the special case when the spin-2 component of torsion vanishes. In this case \Ref{Cll} holds for all null vectors, and torsion-full NGCs coincide with the metric ones. Accordingly, we recover the familiar frame-invariance of shear, twist and expansion, and the same propagation equation, of the metric case. The first property follows from  \Ref{rsLorentz} once we observe that $T_{llm} = \bar T_{llm}\stackrel\nt =0.$
For the second, \Ref{rsrs'} reduces to 
\be\label{primedCll}
\r'\stackrel \nt=\og\r-\f i2\hat T_\m l^\m,\qquad \s'=\og\s, \qquad \t'\stackrel \nt=\og\t+\f13\check T_\m m^\m +\f i2\hat T_\m m^\m,
\ee
which can also be expressed in terms of torsion recalling that $\check T^\m=\check C^\m$ and $\hat T^\m =-2\hat C^\m$.
Hence,
\begin{subequations}\label{Ddec1S}\begin{align}
 D h &\stackrel\nt = \og k h,  \\ 
 D \z &\stackrel\nt = \big(\og\t+\bar{\og\pi} \big) h- \Big(\og\r - \f i2\hat T_\m l^\m\Big)\z- \og\s\bar\z ,
 \end{align}\end{subequations}
which coincide with the purely metric equations (\ref{Dgh},\ref{Dgz}), once we recall that the gauge-condition \Ref{gaugeDm} refers now to the torsion-full connection, 
and
\be
\pi\stackrel\nt =\og\pi -\f13\check T_\m m^\m +\f i2\hat T_\m m^\m, \qquad 2i\im(\eps) \stackrel\nt = 2i\im(\og\eps) + \f i2\hat T_\m l^\m.
\ee
Similar formulas and the same conclusions can be derived for the more general case \Ref{Special}.

\section{Raychaudhuri equation with torsion}

In the light of the relation between $\th'$ and $\th$ of \Ref{th'}, 
the simplest way to derive the Raychaudhuri equation for $\th'$ is to first derive an equation for $D\th$, which can be done following the same procedure of the Levi-Civita case, and then add the extra contribution from the torsion.
We first compute using  \Ref{commRiem}
\be\label{DB}
l^\r\na_\r (g^{\m\n}B_{\m\n}) = -R_{\m\n}(\G)l^\m l^\n - T^\l{}_{\m\n}l^\m  \na_\l l_\n - B_{\m\n} B^{\n\m} +\na_\m(k l^\m).
\ee
Using $g^{\m\n}B_{\m\n} = \na_\m l^\m = \th+k$, where $k$ is now the torsion-full inaffinity, we have
\be
\label{DB1}
D\th = -R_{\m\n}(\G)l^\m l^\n - T^\l{}_{\m\n}l^\m  \na_\l l_\n - B_{\m\n} B^{\n\m} + k^2 + k\th.
\ee
If we want an equation in terms of the frame-invariant quantities defined by $B^\bot$, we can use  \Ref{lB} to derive 
precisely the same relation as in the metric case, 
\be
B^\bot_{\m\n} B^\bot{}^{\n\m} 
= B_{\m\n}\bot^{\m\s}\bot^{\n\r} B_{\r\s} = B_{\m\n} B^{\n\m} -k^2, 
\ee
hence
\be\label{Dth}
D\th = -\f12\th^2  -\s_{\m\n}\s^{\m\n}+\om_{\m\n}\om^{\m\n}-R_{\m\n}(\G)l^\m l^\n - T^\l{}_{\m\n}l^\m  \na_\l l^\n  + k\th.
\ee
However this equation should \emph{not} be taken as the Raychaudhuri equation in the presence of torsion, because $\th$ does not have the geometric interpretation of the expansion of the congruence. 
This was discussed in \cite{Luz:2017ldh} (see also \cite{Dey:2017fld,Puetzfeld:2018cnf}), and starting from the observation that the true displacement tensor is \Ref{B'def}, the expansion was identified with $\th'$.
We have confirmed this by looking at the propagation equations \Ref{Ddec1}. To derive the equation for $\th'$, we rewrite $B_{\m\n}$ in terms of the true displacement tensor $B'{}^\bot{}_{\m\n}$. Using \Ref{lB'} we compute
\begin{align}\nn
B'{}^\bot_{\m\n}B'{}^\bot{}^{\n\m} &= B'_{\m\n}\bot^{\n\r}\bot^{\m\s} B'_{\r\s} = B'_{\m\n} B'{}^{\n\m} - k^2 + 2 T_{\r,\a\m} l^\r l^\a B'{}^{\m\n} n_\n 
+(T_{\m,\n\r} l^\m n^\n l^\r)^2
\\ & = B_{\m\n} B{}^{\n\m} - k^2 + 2 T_{\r,\a\m} l^\a ( l^\r  B{}^{\m\n} n_\n + B^{\m\r}) 
+ 4T_{\m,\n\r} T_{\a,\b\g} l^\n l^\b m^{(\a}\bar m^{\r)} m^{(\g}\bar m^{\m)}, \label{BB'}
\end{align}
where in the last equality we used \Ref{B'def} to substitute $B$ for $B'$, and the fact that
\begin{align}\label{yo}
& T_{\m,\n\r} l^\n T^{\r,\l\m} l_\l   +2T_{\r,\s\m} l^\r l^\s T^{\m,\l\n} l_\l n_\n +(T_{\m,\n\r} l^\m n^\n l^\r)^2 \nn\\ &
= T_{\m,\n\r} T_{\a,\b\g} l^\n l^\b ( g^{\a\r} g^{\g\m}+2g^{\a\r} l^\m n^\g + l^\a n^\r l^\m n^\g) =  4T_{\m,\n\r} T_{\a,\b\g} l^\n l^\b m^{(\a}\bar m^{\r)} m^{(\g}\bar m^{\m)}.
\end{align}
The terms linear in $B$ are clearly a novelty with respect to the standard metric calculation. They could be compactly written as
\be
2 T_{\r,\a\m} l^\a ( l^\r  B{}^{\m\n} n_\n + B^{\m\r} ) = 2 T_{\r,\a\m} l^\a \bot^{\n\r} \na_\n l^\m,
\ee
however this is not useful since the factor 2 in the second term above will cancel with a corresponding term in \Ref{DB1}.
Plugging \Ref{th'} and \Ref{BB'} in \Ref{DB1} we derive
\begin{align}\label{Dth'}
D\th' &=B'{}^\bot_{\m\n}B'{}^\bot{}^{\n\m} + k\th' -R_{\m\n}(\G)l^\m l^\n +2\big(DT_{\m,\l\n}\big)l^\l m^{(\m}\bar m^{\n)}  \\\nn & \qquad + T_{\r,\a\m} l^\a ( 2 l^\r  B{}^{\m\n} n_\n + B^{\m\r} )+  4T_{\m,\n\r} T_{\a,\b\g} l^\n l^\b m^{(\a}\bar m^{\r)} m^{(\g}\bar m^{\m)}.
\end{align}
We now see from \Ref{yo} that replacing $B$ with $B'$ in the linear terms of \Ref{BB'} has the simple effect of replacing the last term of \Ref{BB'} with $T_{lnl}^2$, therefore
\begin{align}\label{Dth'2}
D\th' &= - B'{}^\bot_{\m\n}B'{}^\bot{}^{\n\m}  + k\th' -R_{\m\n}(\G)l^\m l^\n +2\big(DT_{\m,\l\n}\big)l^\l m^{(\m}\bar m^{\n)}  \\\nn & \qquad + T_{\r,\a\m} l^\a ( 2 l^\r  B'{}^{\m\n} n_\n + B'{}^{\m\r} )+  (T_{\m,\n\r} l^\m n^\n l^\r)^2. 
\end{align}
It is not yet in the desired form, as we would like to single out in the right-hand side the quantities describing the geometry of the NGC, like $\r'$ and $\s'$. This is immediately done for the term quadratic in $B$ using \Ref{B2g} as usual.
 However we can expect from the discussion in the previous Sections that it will not be possible to express the linear terms using the orthogonal components alone, but that the parallel ones \Ref{B||} will also appear.
It is in our opinion easiest and geometrically most transparent to work with the NP formalism. Starting from the expression \Ref{Dth'}, we use the spin coefficients to represent the gradient of $l$ (see \Ref{nablal} in the Appendix~\ref{AppNP}), finding
\begin{align}\label{cian1}
& T^{\n,\r\m}l_\r \na_\n l_\m  = \t T_{ll\bar m}-\r T_{ml\bar m}-\s T_{\bar m l\bar m} +{\rm c.c.} \\
& 2 T_{\r,\a\m}  l^\r l^\a  n^\n\na_\n l^\m  = -2 \t T_{ll\bar m}+{\rm c.c.} \label{cian2} \\
& 4T_{\m,\n\r} T_{\a,\b\g} l^\n l^\b m^{(\a}\bar m^{\r)} m^{(\g}\bar m^{\m)} = (T_{ml\bar m})^2 +T_{mlm}T_{\bar m l \bar m} +{\rm c.c.} \label{cian3}
\end{align}
whose sum gives
\be\label{ecco}
-(\t'+T_{mln}) T_{ll\bar m} - \r' T_{ml\bar m} -\s' T_{\bar m l \bar m} +{\rm c.c.}
\ee
in terms of the primed quantities which capture the geometric properties of the torsion-full  NGC.
Using \Ref{ecco} and the usual irrep decomposition \Ref{B2g} for the $(B'{}^\bot)^2$ term we land on the desired result,
\begin{align}\label{RaychaT}
D\th' &= -2\re(\r'{}^2)-2|\s'|^2
+ k\th' -R_{\m\n}(\G)l^\m l^\n +2\big(DT_{\m,\l\n}\big)l^\l m^{(\m}\bar m^{\n)}  \\\nn & \qquad -2\re\Big( (\t'+T_{mln}) T_{ll\bar m} + \r' T_{ml\bar m} + \s' T_{\bar m l \bar m}\Big). 
\end{align}
This is the Raychaudhuri equation for a NGC with arbitrary spacetime torsion.
Notice the explicit presence of the drift term, namely the non-orthogonal component \Ref{B||} for $B'$. 
Even though all geometric quantities are not frame-invariant, but transform like \Ref{rsLorentz} and \Ref{tauI}, the resulting formula is frame-invariant.

Since $\th'=\og\th$, it may be of interest to rewrite the same equation in terms of the Levi-Civita quantities through the relations given by \Ref{rsrs'} and \Ref{tt'}. This can be obtained substituting in \Ref{RaychaT}
\be\label{BBgen}
B'{}^\bot_{\m\n}B'{}^\bot{}^{\n\m} =  \og B^\bot_{\m\n} \og B^\bot{}^{\n\m} -4 i C_{lm\bar m} \, \im(\og\r) +2 (C_{lm\bar m})^2
\ee
and  expressing the curvature using
\Ref{RicciGC}.
We refrain from writing here the resulting expression since no significative simplification occurs. The situation changes when torsion satisfies the special condition \Ref{Cll} or $\bar T=0$.

\subsection*{Spin-2-less torsion}

When $\bar T=0$, it is easy to compute
\begin{align}
& R_{\m\n}(\G)l^\m l^\n \stackrel\nt= R_{\m\n}(g)l^\m l^\n +\f23 (D-k) (\check T_\m l^\m)  -\f12(\hat T_\m l^\m), \\
& B'{}^\bot_{\m\n}B'{}^\bot{}^{\n\m} \stackrel\nt=  \og B^\bot_{\m\n} \og B^\bot{}^{\n\m} - 2\im(\og\r)\, \hat T_\m l^\m-\f12 (\hat T_\m l^\m)^2,\label{BBCll} \\
& 2\re\Big( (\t'+T_{mln}) T_{ll\bar m} + \r' T_{ml\bar m} + \s' T_{\bar m l \bar m}\Big) \stackrel\nt= - \f13 \og\th \, \check T_\m l^\m + 2\im(\og\r)\, \hat T_\m l^\m + (\hat T_\m l^\m)^2.\label{ReAmCll}
\end{align}
We see that various terms cancel out, and using $\th'=\og\th$, \Ref{RaychaT} reduces to 
\begin{align}\label{RaychaS}
D\og\th &\stackrel\nt= - 2\re(\og\r^2) -2|\og\s|^2  -R_{\m\n}(g)l^\m l^\n + (k+\f13\check T_\m l^\m)\og\th.
\end{align}
This is exactly the Raychaudhuri equation for a metric NGC with inaffinity $\og k=k-c$, see \Ref{cnt},
as expected from the discussion below \Ref{Cll}.

\subsection*{Special aligned torsion}
Given the wide utility of the Raychaudhuri equation, let us also provide explicit formulas showing that one recovers exactly the metric one also in the more general case \Ref{Special}, when the spin-2 part of torsion is not completely vanishing. In this case the algebra is a bit more involved but the result the same. We have
\begin{align}\label{RllSpecial}
& R_{\m\n}(\G)l^\m l^\n \stackrel{\Ref{Special}}= R_{\m\n}(g)l^\m l^\n - 2 (D T_{\m,\n\r})m^\m\bar m^\n l^\r - (\r+\bar\r)\bar T_{lln}+(\r-\bar\r)\bar T_{lm\bar m}
\\\nn & \qquad -2\re(\r \bar T_{\bar mlm} +\bar\s T_{mlm}) +T_{lln}(T_{ml\bar m}+T_{\bar mlm}) +C_{ml\bar m}^2+C_{\bar mlm}^2 + 2T_{mlm}T_{\bar ml\bar m}.
\end{align}
In this expression and the following manipulations care is needed to keep track of the full tensor $T$ and its spin-2 part $\bar T$. The decomposition of the various projections are reported in \Ref{Tproj}, in particular $T_{mlm}\equiv\bar T_{mlm}$, and we also notice that
\be\label{CTrel}
C_{ml\bar m}=T_{ml\bar m} +\f32 i \hat T_\m l^\m.
\ee
The $DT$ term in \Ref{RllSpecial} is immediately seen to cancel the corresponding one in \Ref{RaychaT}, but the rest is more tricky. 
The terms linear in $\s$ cancel those in \Ref{RaychaT}, leaving only the squared-torsion contribution 
$2T_{mlm}T_{\bar ml\bar m}$ which cancels out the last term in the second line of \Ref{RllSpecial}.
The terms linear 
in $\r$ coming from the second line of \Ref{RaychaT} minus the second line of \Ref{RllSpecial} give
\be
 - \r T_{ml\bar m} -\bar \r T_{\bar mlm} + \r \bar T_{\bar mlm}+\bar \r T_{ml\bar m} = -(\r+\bar\r)\f13\check T_\m l^\m +(\r-\bar \r)(-\bar T_{lm\bar m}+i\hat T_\m l^m),
\ee
where we used
\be
\bar T_{ml\bar m}-\bar T_{\bar mlm}= \bar T_{lm\bar m}.
\ee
Combining them with the real and imaginary parts of $\r$ coming from (minus) the first line of \Ref{RllSpecial}, we get
\begin{align}
& (\r+\bar\r)(\bar T_{lln} -\f13\check T_\m l^\m) = (\r+\bar\r)T_{lln} = -c \, \og\th+ T_{lln}(T_{ml\bar m}+T_{\bar mlm}),
\end{align}
with the squared-torsion terms cancelling the corresponding ones in the second line of \Ref{RllSpecial}, and
\begin{align}
& (\r-\bar\r)(-2\bar T_{lm\bar m} +i\hat T_\m l^m) = (\og\r-\bar{\og\r}-T_{lm\bar m}) (-2C_{lm\bar m}),
\end{align}
with the $\im(\og\r)$ term cancelling the corresponding one coming from \Ref{BBgen}. After all these cancellations, we are left with
\begin{align}
D\th' &= - \og B^\bot_{\m\n} \og B^\bot{}^{\n\m} + (k-c)\th' -\og R_{ll}  \\\nn & \qquad 
- 2 (C_{lm\bar m})^2+2 T_{lm\bar m} C_{lm\bar m} 
- C_{ml\bar m}^2 - C_{\bar mlm}^2 + T_{ml\bar m}^2 + T_{\bar mlm}^2.
\end{align}
A little algebra using \Ref{CTrel} or the irrep decomposition \Ref{Cirreps} shows that the second line vanishes identically, and from the first line we recover the metric Raychaudhuri equation with $\og k=k-c$, this time $c$ given as in \Ref{Special}.

\section{Optical equations with torsion}

Proceeding like in the metric case, but taking into account torsion through \Ref{commRiem}, the geodesic deviation equation gives
\begin{align}\label{D2eta}
D^2 \eta^\m 
= R^\m{}_{\l\r\n}(\G)l^\l l^\r\eta^\n +\eta^\n\na_\n(k l^\m)+ D(T^\m{}_{\n\r} l^\n \eta^\r),
\end{align}
with the mixed terms $T B$ cancelling out. Using \Ref{etapar}, the gauge choice \Ref{gaugeDm} and projecting along $l$ and $\bar m$ we find\footnote{Projecting along $n$ gives the uninteresting equation for $g$, which we stop writing at this point.}
\begin{subequations}\begin{align}\label{Ddec2}
& D^2 h = 2kDh+hDk+hk^2 + l^\m D\big(T_{\m,\n\l}l^\n \eta^\l\big), \\
& D^2 \z =R_{\m\l\r\n}(\G)m^\m l^\l l^\r\eta^\n + km_\m\eta^\n\na_\n l^\m + D\big(T_{\m,\n\l}m^\m l^\n \eta^\l\big).
\end{align}\end{subequations}
From the $h$ equation (substituiting \Ref{Dh} in \Ref{Ddec2}) we find again an identity, and from the $\zeta$  equation the very same relation \Ref{dev2} as in the metric case, except this time the covariant derivatives and curvature tensor are torsion-full. This similarity is an advantage of working with $B$ at this intermediate stage.
In the next step however, when we use again \Ref{Ddec1} to get rid of the first derivatives of $h$ and $\z$, we introduce an explicit dependence on the torsion field. Recalling that we cannot restrict to $h=0$ since orthogonality is not preserved, we obtain a system of three equations,
\begin{subequations}\label{opt1}\begin{align}
& D \r = \r{}^2 + |\s|^2 +k\r -R_{mll\bar m }(\G) + A_\m \bar m^\m, \\
& D \s = (\r+\bar\r)\s +k\s -R_{mllm }(\G) + A_\m m^\m, \\
& D \t = \r \t +\s\bar{\t} -R_{mlln }(\G)  + A_\m n^\m,
\end{align}\end{subequations}
where we defined
\be
A_\m:=T_{\n,\l\m}(\t l^\n - \r m^\n - \s\bar m^\n)l^\l.
\ee

There are two important differences with the metric case. First, we can not restrict to abreast geodesics, since as we have seen $h=0$ is not conserved along the torsion-full geodesic. Hence, all three equations have to be satisfied. Second, the components of the curvature have additional terms than just the Riemann tensor, as reviewed earlier in \Ref{Riemirreps}. The relevant ones for \Ref{opt1} are
\begin{align}\label{RG1}
& R_{mll\bar m}(\G) = C^{\scr A}_{mll\bar m} - \f12R_{ll} = C^{\scr A}_{mll\bar m} - \Phi_{00}, \\
& R_{mllm }(\G) = C_{mllm} = -\Psi_0, \\
& R_{mlln}(\G) = C_{mlln} + C^{\scr A}_{mlln} -\f12R_{ml} = - \Psi_1-\Phi_{01}+ C^{\scr A}_{mlln} + \f12R^{\scr A}_{lm}.
\end{align}

Finally, re-expressing the spin coefficients in terms of the geometric primed coefficients through \Ref{rsrs'}, we arrive at
\begin{subequations}\label{opt'}\begin{align}
& D \r' = \r'{}^2 + |\s'|^2 +k\r' +\Phi_{00} -C^{\scr A}_{mll\bar m}(\G) - (DT_{\m,\n\r}) m^\m l^\n \bar m^\r \nn\\ &\quad\qquad + \t' T_{ll\bar m} + \r' T_{ml\bar m} + \bar\s' T_{mlm} + T_{mln} T_{ll\bar m}, \\
& D \s' 
= (\r'+\bar\r')\s' +k\s' +\Psi_0 + \t' T_{ll\bar m} - (DT_{\m,\n\r}) m^\m l^\n  m^\r  \nn\\ &\qquad\quad + \bar\r' T_{ml m }+ \s' T_{ml\bar m} +T_{mln} T_{llm}, \\
& D \t' = \r' \t' +\s' \bar{\t}' + \Psi_1+\Phi_{01}- C^{\scr A}_{mlln } - (DT_{\m,\n\r}) m^\m l^\n  n^\r 
 \nn\\ &\qquad\quad + \t'(T_{lln}+ T_{ml\bar m}) + \bar{\t}' T_{mlm} + T_{mln} T_{lln} .
\end{align}\end{subequations}
We can also use this system to rederive the Raychaudhuri equation, 
\begin{align}
D\th' & = -2D\re(\r') \\\nn &= - \r'{}^2 - \bar\r'{}^2 -2 |\s'|^2 +k\th' -2\Phi_{00} +2(DT_{\m,\n\r})l^\n m^{(\m}\bar m^{\n)} 
\\\nn &\qquad 
-2\re\Big( (\t'+T_{mln}) T_{ll\bar m} + \r' T_{ml\bar m} + \s' T_{\bar m l \bar m}\Big),
\end{align}
which coincides with \Ref{Dth'} derived earlier. Notice in particular that the non-Riemannian part $C^{\scr A}_{mll\bar m}(\G) $ disappears from the Raychaudhuri equation because of its antisymmetry.

The optical equations \Ref{opt'} in the presence of torsion are the main result of this paper. 
All quantities, spin coefficients and curvature scalars, contain torsion, and we notice the presence of non-Riemannian components of the curvature.
We also remark that even thought the shear, twist, expansion and drift are explicitly $n$-dependent in the torsion-full case, the optical equations are invariant under the freedom of changing $n$ while keeping $l$ fixed, namely under class-$I$ Lorentz transformations of the adapted tetrad.

\subsection*{Spin-2-less torsion}
As before, we conclude the Section proving equivalence with the metric case for the special case with no spin-2 component of  torsion. Starting from \Ref{RiemmG} we compute
\begin{align}
R_{mll\bar m}(\G) & \stackrel\nt= 
R_{mll\bar m}(g) -\f13 (D-k) \check T_\m l^\m +\f14(\hat T_\m l^\m)^2 +\f i2(D-k)\hat T_\m l^\m, \\
R_{mllm}(\G) & \stackrel\nt= R_{mllm}(g), \\
R_{mlln}(\G) & \stackrel\nt= R_{mlln}(g) -\f13 D\check T_\m m^\m +\f i2 D\hat T_\m m^\m 
+\f i3 \check T_\m l^\m \,\hat T_\n m^\n - \f i6 \check T_\m m^\m \,\hat T_\n l^\n +\f14 \hat T_\m l^\m \,\hat T_\n m^\n. 
\end{align}
Using these together with \Ref{rsrs'}, \Ref{tt'} and the torsional projections listed in  \Ref{Tproj} in the Appendix, \Ref{opt'} reduce to
\begin{align}\label{opt'Cll}
& D \og\r \stackrel\nt= \og\r{}^2 + |\og\s|^2 +\og k\og\r +\og{\Phi}_{00} + \f i6 \check T_\m l^\m \, \hat T_\m l^\m, \\
& D \og\s \stackrel\nt= 2\re(\og\r)\og\s +\og k\og\s +\og{\Psi}_0 , \\
& D \og\t \stackrel\nt=  \og\r\og\t +\og\s \bar{\og\t} + \og{\Psi}_1+\og{\Phi}_{01} +\og\r \Big(\f13\check T_\m m^\m + \f i2 \hat T_\m m^\m\Big)
+\og\s \Big(\f13\check T_\m \bar m^\m - \f i2 \hat T_\m \bar m^\m\Big) 
 \nn\\ &\qquad\quad - i\hat T_\m l^\m \Big(\og\t + \f 13 \check T_\n m^\n +\f i2\hat T_\n m^\n\Big).
\end{align}
We see that we recover the same metric equation for the expansion (as already proved in Section 5) as well as for the shear. The equations for the twist has an additional term, which has a gauge interpretation: Since we have imposed
\be
D m^\m = 0 = \og D m^\m - \Big(\f 13 \check T_\n m^\n - \f i2\hat T_\n m^\n\Big) l^\m +\f i2\hat T_\n l^\n  m^\m,
\ee
there are additional drift and twist contributions introduced by the non-parallel transport of the complex dyad with respect to the Levi-Civita connection.
The last term in the twist equation can in fact be interpreted as
\be
\f i6 \check T_\m l^\m \, \hat T_\m l^\m = 2i (k-\og k) \, \im(\og\eps-\eps)
\ee

Finally we notice that for a completely antisymmetric torsion, the equations for shear and twist match the metric ones, as expected from the fact that the geodesic equations completely coincide. The optical equation for the drift term still differs on the other hand, since this depends on the (non-geodetic) evolution of $n$ as well, which feels even a completely antisymmetric torsion.

\section{Comments and conclusions}
 
In this paper we derived the optical equations for NGCs in the presence of torsion, extending previous results in the literature on the Raychaudhuri equation. Unlike the Raychaudhuri equation, the full set depends also on non-Riemannian components of the curvature. We further noticed that one must include the evolution of the drift term, because along a torsion-full geodesic orthogonality of the connecting vector is not preserved, and thus one cannot restrict attention to abreast bundles. 
Deriving this result provided us with the opportunity to review some less familiar aspects of metric NGCs, and the utility of the NP formalism to study them. 

It is well-known that for completely antisymmetric torsion, the geodesic equation coincides with the metric one. A characteristic of null geodesic congruences, unlike time-like ones, is that also the trace-part of torsion only contributes to an inaffinity difference, without changing the direction of the metric geodesics. 
Accordingly, we have provided explicit formulas for the case of spin-2-less torsion, and showed that the Ryachaudhuri and optical equation for the shear reduce exactly to the metric ones. The one for the twist does not, somehow unexpectedly, but the difference stems only from a gauge condition on the parallel transport of the space-like dyad. Finally the equation for the drift is always different since it depends on non-geodetic evolution. Similar consideration apply to the more general case of spin-2 torsion aligned with the null geodesic vector. 

Having established these equations, future work could explore their explicit solutions and the structure of the torsion components entering, and their relation to the Noether identities and field equations of the specific theory considered.
It would be also interesting to further elaborate on the geometric possibility of the alternative Lie dragging \Ref{Lie2}. Finally, even though torsion-full geodesics do not arise from the conservation law of test matter in Einstein-Cartan of Poincar\'e gauge theory of gravity \cite{Hehl:2013qga}, it could be interesting to explore what happens with the conserved energy momentum tensor used for instance in \cite{DeLorenzo:2018odq}.

\subsection*{Acknowledgments}
We would like to thank Rafael Sorkin and Friedrich Hehl for discussions.

\appendix
\setcounter{equation}{0}
\renewcommand{\theequation}{\Alph{section}.\arabic{equation}}

\section{Newman-Penrose notation}\label{AppNP}
For the tetrad derivatives we have
\begin{align}
& D = l^\m\na_\m, \qquad \D = n^\m\na_\m ,\qquad \d =m^\m\na_\m , \qquad \bar\d =m^\m\na_\m.
\end{align}
For the spin coefficients and curvature scalars we use the standard notation consistent with mostly plus signature, which carries an opposite sign as to the notation with mostly minus signature, see e.g. the Appendix of \cite{Ashtekar:2000hw}. The connection components are represented by twelve complex scalars,
\begin{align}
&  {\a} := -\f12 (n^\m \bar\d l_\m + m^\m \bar\d \bar m_\m)   \qquad  {\b} := -\f12 (n^\m \d l_\m + m^\m \d \bar m_\m) \\
&  {\g} := - \f12 (n^\m \Delta l_\m + m^\m \Delta \bar m_\m)  \qquad  {\eps} := -\f12 (n^\m D l_\m + m^\m D \bar m_\m)  \\
&  {\kappa} := -m^\m D l_\m \qquad  {\tau} := -m^\m \Delta l_\m \qquad { \s} := -m^\m \d l_\m \qquad  {\r} := -m^\m \bar\d l_\m \\
&   { \pi} := \bar m^\m D n_\m  \qquad {\n } := \bar m^\m \Delta n_\m  \qquad { \l} := \bar m^\m \bar\d n_\m \qquad {\m} := \bar m^\m \d n_\m 
\end{align}
whose geometric interpretation is as follows:
$\k$ measures the (orthogonal) acceleration of $l$ (hence it vanishes when $l$ is geodesic), and $2\re(\eps)$ its parallel acceleration (or the inaffinity);  $\s$ its shear and $\r$ its expansion and twist. $\im(\eps)$ is the twisting of $(m,\bar m)$ in $S$ while transported along $l$, and $\pi$ its component along $n$. Finally
$\bar\a-\b=-\bar m^\m\d m_\m$ is the 2d connection coefficient.
The corresponding quantities for $n$ are $\n$, $2\re(\g)$, $\l$ and $\m$, $\im(\g)$ and $\tau$.

In terms of these coefficients we have the general decomposition
\be\label{nablal}
\na_\m l_\n = -\eps n_\m l_\n +\k n_\m \bar m_\n- \g l_\m l_\n + (\bar \a +\b)\bar m_\m l_\n +\t l_\m \bar m_\n-\s \bar m_\m \bar m_\n -\r m_\m \bar m_\n +{\rm cc},
\ee
which is used in the main text to derive (\ref{cian1}--\ref{cian3}).

If $l$ is geodesic, then 
\be
\k=0, \qquad \eps+\bar\eps=k, \qquad D n^\m=-k n^\m, \qquad Dm^\m = \bar\pi l^\m +(\eps-\bar\eps)m^\m.
\ee
If it is furthermore hypersurface orthogonal, e.g. $l_\m = N \p_\m \Phi$, then $\im(\r)=0$, and if $N=1$ then $\t = \bar\a+\b$.

For the curvature components, we have
\begin{align}\label{defPsi}
& \Psi_0:= C_{lmlm}, \quad \Psi_1:= C_{lnlm}, \quad \Psi_2:=-C_{lmn\bar m}=\f12(C_{lnln}-C_{lnm\bar m}),\\ 
& \Psi_3:=-C_{lnn\bar m}, \quad \Psi_4:=C_{n\bar mn \bar m}
\end{align}
and
\begin{align}\label{defPhi}
& \Phi_{00}:= \f12 R_{ll}, \quad \Phi_{01}:= \f12 R_{lm}, \quad \Phi_{10}:= \f12 R_{l\bar m}, \quad \Phi_{02}:= \f12 R_{mm}, \quad  \Phi_{20}:= \f12 R_{\bar m\bar m}, \\
& \Phi_{12}:= \f12 R_{nm}, \quad \Phi_{22}:= \f12 R_{nn}, \quad \Phi_{21}:= \f12 R_{n\bar m}, \quad \Phi_{11}:= \f14 R_{ln}+\f14 R_{m\bar m}.
\end{align}

Notice that the above quantities, introduced by Newman and Penrose for the Levi-Civita connection and Riemann curvature, can be immediately extended to a connection and curvature with torsion, and in this in this sense that they are used in the present paper. In that case there are also additional components to the curvature than \Ref{defPsi} and \Ref{defPhi}, see \Ref{Riemirreps}. For these, as well as for torsion itself, we are not aware of a consensual NP notation. We refrain from investigating the issue in details here, as it would go beyond the scope of this paper. We merely point out the contributions of the three irreps to the various (complex) projections, since this was used in the main text.
With the following convention for the area 2-form on $S$ and the NP tetrad determinant,
\be
^{(2)}\eps:=im\w\bar m, \qquad ^{(2)}\eps_{\m\n}:=2im_{[\m}\bar m_{\n]}, \qquad i \eps_{\m\n\r\s}l^\m n^\n m^\r \bar m^\s=1,
\ee
where $\eps_{\m\n\r\s}$ are the components of the spacetime volume form (with conventions $\eps_{0123}=\sqrt{-g}$), 
we have
\be
\eps_{\m\n\r\s}l^\m n^\n m^\r = -im_\s, \qquad \eps_{\m\n\r\s}l^\m m^\n \bar m^\r = i l_\s, \qquad \eps_{\m\n\r\s}n^\m m^\n \bar m^\r = - i n_\s,
\ee
and
\begin{subequations}\label{Tproj}\begin{align}
& T_{lln} = \bar T_{lln} - \f13 \check T\cdot l, \qquad T_{llm} = \bar T_{llm}, \qquad T_{lnm} = \bar T_{lnm}+ \f13 \check T\cdot m -i\hat T\cdot m,  \\
& T_{nnl} = \bar T_{nnl} - \f13 \check T\cdot n, \qquad T_{nnm} = \bar T_{nnm},\qquad T_{nlm} = \bar T_{nlm} + \f13 \check T\cdot m + i\hat T\cdot m, \\
& T_{lm\bar m} =\bar T_{lm\bar m} +i \hat T\cdot l,\qquad T_{nm\bar m} = \bar T_{nm\bar m} -i\hat T\cdot n, \\
& T_{mln}=\bar T_{mln} -i \hat T\cdot m, \qquad T_{mm\bar m}=\bar T_{mm\bar m}+\f13\check T\cdot m, \\
& T_{mlm}=\bar T_{mlm}, \qquad T_{mnm}=\bar T_{mnm}, \qquad  T_{mml} = \bar T_{mml}, \qquad T_{mmn} = \bar T_{mmn}, \\
& T_{ml\bar m}=\bar T_{ml\bar m} +\f13\check T\cdot l -i\hat T\cdot l, 
\qquad T_{mn\bar m}=\bar T_{mn\bar m}+\f13\check T\cdot n+i\hat T\cdot n.
\end{align}\end{subequations}

\providecommand{\href}[2]{#2}\begingroup\raggedright\endgroup

\end{document}